# BINARY PULSAR SHOCK EMISSIONS AS GALACTIC GAMMA-RAY SOURCES

MARCO TAVANI

*Joseph Henry Laboratories and Department of Physics*
*Princeton University, Princeton, NJ 08544 (USA)*

# BINARY PULSAR SHOCK EMISSIONS AS GALACTIC GAMMA-RAY SOURCES

MARCO TAVANI
*Joseph Henry Laboratories and Department of Physics*
*Princeton University, Princeton, NJ 08544 (USA)*

ABSTRACT. We address several issues regarding the interpretation of galactic $\gamma$-ray sources. We consider powerful pulsars in binaries producing X-ray and gamma-ray *unpulsed* emission from the shock interaction of relativistic pulsar winds with circumbinary material. Nebular mass outflows from companion stars of binary pulsars can provide the right *calorimeters* to transform a fraction of the electromagnetic and kinetic energy of pulsar winds into high energy radiation. We discuss the physics of interaction of relativistic pulsar winds with gaseous material and show that the conditions in pulsar binary systems might be ideal to constrain shock acceleration mechanisms and pulsar wind composition and structure. We briefly discuss the example of the 47 ms pulsar PSR 1259-63 orbiting around a massive Be star companion and monitored by X-ray and gamma-ray instruments during its recent periastron passage. In addition to young pulsars in massive binaries, also a class of recycled millisecond pulsars in low-mass binaries can be interesting high energy emitters.

## 1. Introduction

Recent data of the *EGRET* instrument on board of the *Compton* Gamma-Ray Observatory (CGRO) revealed at least $\sim 40$ point-like $\gamma$-ray sources of low galactic latitude $b_{II} \lesssim 10°$ (Fichtel *et al.*, 1994). *EGRET* confirms several of the previously discovered *COS-B* sources (Swanenburg *et al.*, 1981; Bignami & Hermsen, 1983) and adds a significant number of new ones. Gamma-ray emission (from $\sim$MeV to tens of GeV) is a manifestation of particle acceleration and non-thermal behavior of astrophysical systems. Candidate systems for high energy emission include:
(1) isolated pulsars,
(2) supernova remnants,
(3) pulsars moving in the interstellar medium;
(4) compact stars (e.g., pulsars) with relativistic winds in binaries,
(5) massive binaries with colliding non-relativistic mass outflows,
(6) accreting compact stars.

Revealing the nature of unidentified gamma-ray sources is difficult, and requires monitoring and multiwavelength observations. For example, the $\gamma$-ray source *2CG 195+04 (Geminga)* (e.g., Bignami & Hermsen, 1983), was recently identified as a 237 ms $\gamma$-ray pulsar (Bertsch *et al.*, 1992) only after a crucial *ROSAT* observation (Halpern & Holt, 1992). Accreting sources are not strong $\gamma$-ray emitters (e.g., the black hole candidate *Cyg X-3* is marginally detected by *COMPTEL* up to $\sim$10 MeV, e.g., Hermsen, 1994), a fact related to the difficulty of sustaining non-thermal acceleration and radiative processes in accretion disks. The environment surrounding compact stars must therefore be relatively 'clean' to allow the formation and propagation of $\gamma$-rays. Without invoking the existence of black hole related phenomena, *pulsar magnetospheres* (of $\gamma$-ray pulsars) and *nebular shock environments* are systems considered in a 'conservative' approach to the interpretation of galactic $\gamma$-ray sources. We mainly discuss here high energy shock emission in binary systems containing rapidly rotating pulsars interacting with circumbinary material. The high energy emission might be modulated with the underlying orbital period. These systems (both low-mass and high-mass binaries) are interesting candidates for the low-galactic latitude sources of Fichtel *et al.*(1994).

We address a few topics related to the interpretation of galactic $\gamma$-ray sources in term of a pulsar-driven model of relativistic nebular emission. We do not discuss here isolated $\gamma$-ray pulsars (which are an important class of steady $\gamma$-ray sources, e.g., Thompson, 1994), pulsars moving in the interstellar medium (e.g., PSR 1957+20, Fruchter *et al.*, 1988; Arons & Tavani, 1993, hereafter AT93), OB associations (e.g., Montmerle, 1979), and massive binaries with colliding non-relativistic mass outflows (e.g., the Wolf-Rayet system *WR 140*, Williams *et al.*, 1990; Eichler & Usov, 1993). The physics of interaction of relativistic winds with nebular material is of general interest, and it may also be relevant for transient high energy emission of $\gamma$-ray bursts (Mészàros & Rees, 1992) and soft gamma-ray repeaters (Tavani, 1994d).

## 2. High Energy Emission from Plerions: the Crab Nebula

Pulsars lose their spindown energy mainly by the emission of electromagnetic waves and kinetic energy of relativistic particles. In the absence of a relatively dense environment, these relativistic outflows are practically unobservable, as demonstrated by the lack of detectable high energy emission from the majority of pulsars. However, in a few cases, either a supernova remnant, a dense interstellar medium, or a dense circumbinary material, can provide the right nebular environment to convert a fraction of pulsar spindown energy into observable high energy emission. The nebula acts as a *calorimeter* with conversion efficiency $\varepsilon_a$ into high energy emission in the range $0.01 \lesssim \varepsilon_a \lesssim 0.1$.

Until recently, only nebular environments of pulsars embedded in supernova remnants were available to study. The Crab Nebula is the prototypical example of an X-ray and $\gamma$-ray radiating pulsar-driven nebula (plerion). The Crab Nebula shows *unpulsed* emission from radio to $\gamma$-ray wavelengths[1]. The 33 ms Crab pulsar is embedded in a quasi-spherical 'pulsar cavity' with its relativistic particle and electromagnetic wind confined by a MHD shock at the distance[2] from the pulsar $r_s \sim 3 \cdot 10^{17}$ cm (Rees & Gunn, 1974; Kennel & Coroniti, 1984, hereafter KC84). Observations show that the Crab nebula component dominates (by a factor $\sim 5$) the Crab pulsar component of the spectrum up to a few MeV with photon index $\alpha_1 \sim 2.2$ (e.g., Bartlett *et al.*, 1994). The nebular becomes softer (and subdominant with respect to the pulsar emission) with $\alpha_2 \sim 2.9$ between a few MeV and a few hundreds of MeV (Nolan *et al.*1993). At $\sim 500$ MeV a change in spectral index was observed by *EGRET* up to $\sim 30$ GeV with $\alpha_3 \sim 2.7$ (Nolan *et al.*, 1993, De Jager *et al.*, 1994). The photon index of the Crab nebula at the highest *EGRET* energies is consistent with the observed level of unpulsed TeV emission (Vacanti *et al.*, 1991).

High energy emission of the Crab Nebula is a manifestation of a relativistic pulsar wind originating from the pulsar (Rees & Gunn, 1974). As the pulsar wind (a mixture of electromagnetic fields and particles) interacts with a quasi-static MHD shock at $r_s$, a redistribution in momentum space of the downstream particle distribution occurs and non-thermal synchrotron and inverse Compton (IC) radiation can be emitted[3]. The volume of the emitting region is larger than the inner pulsar cavity, and both particles and fields diffuse out from the central region into the main body of the nebula. The Crab Nebula is the most thoroughly studied plerion and any model of the Crab nebular shock emission is a natural candidate for the study of pulsar shock emission of other systems. In this section, we use the Crab nebular model to introduce the main relevant properties of pulsar winds and nebular emission that we will apply below to binary systems.

2.1 RELATIVISTIC PULSAR WINDS

Pulsar magnetospheres are powerful particle accelerators and a rapidly rotating pulsars as the Crab is capable to emit a powerful wind from the open field line regions. The accelerating voltage drop above a neutron star polar cap region can be very high. For the Crab pulsar is $\Phi_c \sim \Omega^2 \mu/c^2 \sim 10^{13} \mu_{30}/P^2$ Volt, with $\Omega = 2\pi/P$ the spin frequency of a pulsar of period $P$, and $\mu = \mu_{30} \, 10^{30}$ (in c.g.s. units) the neutron star magnetic moment. Acceleration and $e^{\pm}$-pair creation inside the light cylinder radius ($R_{lc} = c/\Omega \sim 1.57 \cdot 10^8$ cm for the Crab pulsar) leads to the formation of a relativistic particle wind. At sufficiently large distance from the pulsar, the magnetic field $B_1$ is mostly toroidal and the MHD approximation can be applied to the pulsar wind. The wind has a flow rate in ions $\dot{N}_i$ and $e^{\pm}$-pairs $\dot{N}_{\pm}$, a flow Lorentz factor $\gamma_1$, and a ratio of the electromagnetic energy flux (Poynting flux) to kinetic energy flux upstream of the shock, $\sigma \equiv B_1^2/4\pi[N_i m_i + m_{\pm}(N_+ + N_-)]\gamma_1 c^2$. A MHD model of a pulsar wind can therefore be characterized by a set of five parameters

$$(\sigma, \gamma_1, B_1, \rho_i/\rho_{\pm}, Z/A) \qquad (1)$$

where we defined the ions' density $\rho_i$, the $e^{\pm}$-pairs' density $\rho_{\pm}$, and the charge to mass ratio for ions $Z/A$.

---

[1] For a summary, see Kennel & Coroniti (1984) and references therein.

[2] The MHD inner shock radius of the Crab Nebula has an apparent size of $10''$ at the distance of 2 kpc. For all practical purposes, pulsar cavities of known systems appear as point sources to X-ray and $\gamma$-ray instruments.

[3] See however Cheung & Cheng (1993) for an alternative model.

The details of acceleration, particle composition and electromagnetic structure of relativistic winds near $R_{lc}$ are poorly constrained at present. Qualitatively, the quantity $\sigma$ is expected to be relatively large near the pulsar, i.e., $\sigma \sim 10^4$, with a rate of pair creation $\dot{N}_{\pm} \sim 10^{38}\,\text{s}^{-1}$ and Lorentz factor $\gamma_i \sim 200$ (Coroniti, 1990). As the wind expands out from the pulsar, it evolves from a high-$\sigma$ to a low-$\sigma$ configuration and the $\gamma$-factor increases. No direct information is available to probe the inner region of a pulsar wind near $R_{lc}$. However, a region very far from $R_{lc}$ (with $r_s/R_{lc} \sim 10^8 - 10^9$ in the case of the Crab Nebula) can be (indirectly) probed by the properties of a MHD shock. An important result is obtained by combining the observational results of the Crab nebula with a detailed hydrodynamic and radiative model of the shock. With the *assumption* of a downstream power-law $e^{\pm}$-pair momentum distribution, the MHD model of the pulsar wind in the Crab Nebula gives $\sigma \sim 0.005$ and $\gamma_1 \gtrsim 10^6$ (KC84). Notice that these values are characteristic of the Crab pulsar wind *only for a very large distance from $R_{lc}$*. The determination of the ions' content of the Crab pulsar wind is more controversial and dependent on the shock acceleration mechanism used to model the Crab perpendicular shock[4]. A particle-in-cell numerical simulation of a perpendicular shock for a pure $e^{\pm}$-pair pulsar wind fails to produce a non-thermal downstream particle distribution (Hoshino *et al.*, 1992, hereafter HAGL92; Gallant & Arons, 1994). A non-thermal downstream momentum distribution of $e^{\pm}$-pairs develops only with the inclusion of a heavy ion component of the pulsar wind dominating the particle kinetic energy flux (HAGL92). Despite the one-dimensional character of these simulations and the approximations of the calculations of HAGL92, the possible presence of ions in the pulsar wind should be considered as an interesting possibility that can be tested by observations.

2.2 NEBULAR SHOCK EMISSION

The high energy emission produced by the pulsar wind termination shock of the Crab Nebula depends on shock acceleration mechanism. The fundamental fact about Crab-like pulsar-driven shocks is that the geometry of the shock for a spherically symmetric outflow is typical of a 'perpendicular shock' rather than of a 'parallel shock'.

---

[4] A perpendicular shock is characterized by a shock surface whose normal vector is perpendicular to the upstream magnetic field, as expected in the idealized model of the Crab pulsar wind with a toroidal magnetic field confined by a spherically symmetric inner cavity.

The particle acceleration mechanism is therefore different from the (first-order) diffusive Fermi acceleration mechanism that is applicable to 'parallel shocks' (e.g., Bell, 1978; Blandford & Ostriker, 1978; Jones *et al.*, 1994). Shock drift acceleration models of perpendicular shocks might explain pure $e^{\pm}$-pairs acceleration in the Crab Nebula, but only if cross field diffusion is present (e.g., Jones *et al.*, 1994). A developed model of perpendicular shocks currently available is the model of HAGL92, that we discuss here. In order to explain the properties of the Crab Nebula, it is necessary to identify a shock acceleration mechanism able to transform a relativistic Maxwellian particle distribution upstream of the shock into a downstream power-law distribution of $e^{\pm}$-pairs. The downstream distribution of $e^{\pm}$-pairs per energy interval becomes, $N(\gamma)\,d\gamma \sim \gamma^{-s}$, with $s$ an exponent. Simulations indicate that a resonant mechanism of $e^{\pm}$-pair acceleration mediated by ions' gyromotion satisfies the necessary requirements for the Crab Nebula (HAGL92).

Once a downstream non-thermal particle distribution is established, radiation occurs by synchrotron emission and inverse Compton scattering. The magnetic field of the Crab Nebula upstream of the inner shock is

$$B_1(r_s) = \left(\frac{\sigma}{1+\sigma}\right)^{1/2} \left(\frac{\dot{E}_R}{r_s^2 c f_p}\right)^{1/2} \simeq 3 \cdot 10^{-5} \left[\frac{\sigma}{0.005} \frac{n_2 v_7^2}{(1+\sigma)}\right]^{1/2} \text{ G} \qquad (2)$$

with $f_p$ the fraction of the total solid angle for the pulsar wind expansion ($f_p = 1$ for isotropic expansion). In Eq. (2) we used a typical combination of density and gas velocity at the Crab Nebula shock, $n_2 = n(r_s)/(10^2 \text{ cm}^{-3})$ and $v_7 = v(r_s)/(10^7 \text{ cm s}^{-1})$. Note that $B_1(r_s)$ depends ultimately on the value of the gas pressure at $r_s$ and, to first order, is independent of the pulsar spindown power. In the case of the Crab Nebula, the nebular pulsar pressure is calculated by taking into account the energy accumulation during the pulsar lifetime (Rees & Gunn, 1974). The Rankine-Hugoniot equations of a relativistic shock give for a low-$\sigma$ pulsar wind the value of the downstream magnetic field $B_2 = 3\,B_1$ (KC84). The downstream magnetic field $B_2$ enters in the calculation of the shock synchrotron emissivity.

The observed efficiency of conversion of pulsar spindown energy into nebular high energy emission is near $\sim 10\%$ (KC84), and the numerical model of HAGL92 is in agreement with this relatively high efficiency. The high energy spectrum of the Crab Nebula is explained in terms of synchrotron radiating $e^{\pm}$-pairs being accelerated at the inner shock and progressively diffusing out in the central nebula (KC84). Inverse Compton scattering against a plerionic diffuse low-energy photon background may play a role for the nebular emission above $\sim 500$ MeV (De Jager *et al.*, 1994).

2.2 A *GEDANKEN* EXPERIMENT: COMPRESSING PULSAR-DRIVEN CAVITIES

A pulsar cavity in a supernova remnant is by definition quasi-static without appreciable time variability. If pulsars in supernova remnants were the only systems available to study pulsar winds, there would be no hope to see how a MHD shock changes its radiative properties as the pulsar cavity changes in size. However, one can think of a *gedanken* experiment with pulsar cavities. Let us imagine a rapidly rotating and energetic pulsar at the center of a cavity in a plerionic and calorimetric nebula. The nebular high energy emission depends on the pulsar wind parameters of Eq. (1). Given the relatively large number of (model-dependent) parameters, it might be difficult to obtain a unique model of emission. Now, let us imagine to *compress* progressively the pulsar cavity (e.g, by increasing the gas pressure at the MHD shock) from a large radius with $r_s/R_{lc} \sim 10^9$ as in case of the Crab Nebula to a much smaller radius. The local magnetic field strength varies, and the corresponding change of the high energy intensity and spectrum will follow a pattern proper to a particular set of pulsar wind parameters. For example, the magnetosonic $e^\pm$-pair acceleration mechanism for a perpendicular shock mediated by heavy ions can be tested for $r_s$ comparable or less[5] than the ions' Larmor radius, $R_{L,i} \sim 3 \cdot 10^{17} \, (A/Z) \, (\gamma_1/10^6)(B_1/10^{-5} \, \text{G})^{-1}$. A transition to a power-law high energy emission to a quasi-thermal spectrum of emission is expected in the model of HAGL92 as the shock radius decreases and the magnetosonic mechanism becomes progressively more inefficient for $R_{L,i} \gtrsim r_s$. Analogously, a transition from a very-low-$\sigma$ to a moderately-low-$\sigma$ pulsar wind might in principle produce observable effects, as $r_s$ decreases. Therefore, *time dependent* nebular environments can provide ideal calorimeters to test theories of shock acceleration and pulsar wind composition.

Obviously, it is not possible to perform this *gedanken* experiment with plerionic supernova remnants. However, pulsars in binary systems can be in the right environment to produce time variable nebular emission.

## 3. Pulsar-driven Nebular Emission of Binary Pulsars

Binary pulsars are interesting systems which may provide environments where we can test the pulsar wind model and acceleration mechanism established for the Crab Nebula. There is no reason in principle to assume that the pulsar wind parameters of Eq. (1) are the same for all pulsars. In particular, binary pulsars are characterized by a ratio $r_s/R_{lc}$ different from the Crab value.

### 3.1 PULSAR-DRIVEN SHOCKS IN CIRCUMBINARY MATERIAL

We consider here a binary containing a pulsar interacting with a mass outflow (gaseous non-relativistic wind from the companion star surface). The discussion presented here can be applied to high-mass and low-mass systems with proper modifications. We follow here the discussion of Tavani, Arons & Kaspi (1994; hereafter TAK94).

The mass outflow has in general a density profile $\rho(R) = \rho_0 (R/R_*)^{-n'}$, where $R_*$ is the radius of the star, $R$ the distance from its surface, and $\rho_0 = \dot{M}/[4\pi f R_*^2 v_0]$, with $\dot{M}$ the mass outflow rate, $v_0$ the outflow velocity at the surface of the companion star, and $f$ the fraction of $4\pi$ steradians into which the mass loss occurs. Typically, the surface density $\rho_0$ is of the order of $10^{-12} - 10^{-11}$ g cm$^{-3}$, for both massive Be-star-like companions and low-mass companions. For a constant velocity mass outflow (typical of low-mass star outflows), the 'outflow exponent' is $n' = 2$. More generally, for massive companions, the outflow exponent is in the range $2 < n' < 4$ (Waters *et al.*, 1988). The mass outflow pressure at distance $R$ from the surface of the companion star has a radial dependence of the form, $P(R) = \rho(R) v(R)^2 \sim (R/R_*)^{n'-4}$, obtained with the use of the continuity equation in spherical symmetry. Note that the outflow velocity changes with distance as $v(R) \sim (R/R_*)^{n'-2}$, i.e., it increases for $n' > 2$ as suggested by observations of Be mass outflows (Waters *et al.*, 1988). Near the pulsar orbit, the outflow velocity can reach values of order of 100 km s$^{-1}$. We define the mass outflow parameter[6], $\Upsilon = \dot{M}_{-8} v_{0,6}$, where $\dot{M} = (10^{-8} M_\odot$ yr$^{-1}) \dot{M}_{-8}$, and $v_0 = (10^6$ cm s$^{-1}) v_{0,6}$. It is useful to parametrize the properties of the companion star mass outflow by specifying the parameters $(\Upsilon, n')$.

---

[5] We notice that for a radial dependence of electromagnetic pulsar waves $B_1 \sim r^{-1}$, the ratio $r_s/R_{L,i}$ varies only because of a change of $\sigma$ with radius.

The radial distance $r_s$ of the apex of the shock region from the pulsar is obtained by equating the pulsar radiation pressure with the ram pressure of the mass outflow (e.g., AT93). The radius $r_s$ is the solution of the equation (TAK94),

$$\frac{r_s/d(\theta)}{[1-(r_s/d(\theta))]^{2-n'/2}} = \left(\frac{\dot{E}_R f}{c \dot{M} v_0}\right)^{1/2} \left(\frac{R_*}{d(\theta)}\right)^{n'/2-1} \simeq \frac{7.2 f^{1/2}}{\Upsilon^{1/2}} \left(\frac{R_*}{d(\theta)}\right)^{n'/2-1}, \quad (3)$$

where we have assumed $R_* \ll d$, with $d$ the pulsar-companion star separation. In general, $r_s$ is a function of the true anomaly $\theta$, $n'$, and $\Upsilon$. Pulsar radiation pressure of a rapidly rotating pulsar can resist the compressing tendency of mass outflows in binaries and avoid accretion (e.g., Tavani, 1991). This conclusion follows from the fact that *for high-mass binaries* the gravitational accretion radius

$$R_G \equiv 2GM_p/v_*^2 \approx (3 \cdot 10^{12} \text{ cm}) M_{p,1.4}/v_{*,7}^2 \quad (4)$$

(with $M_{p,1.4}$ the pulsar mass in units of 1.4 $M_\odot$ and $v_{*,7} = v_*/(10^7$ cm s$^{-1})$, where $v_* = \sqrt{v^2 + v_{orb}^2}$, and $v_{orb}$ is the orbital velocity) is small[7] compared to the shock radius $r_s$ for a large range of outflow parameters $\Upsilon$ and $n'$. However, the effects of radio pulse dispersion and absorption as the pulsar passes into the densest part of the outflow may suppress the pulsed radio signal. The pulsar can be totally or partially 'hidden' or enshrouded during its orbital motion.

The magnetic field $B_1(r_s)$ at the apex of a pulsar cavity in a typical binary is given by Eq. (2) with $n_2(r_s) \sim 10^6 - 10^8$.

---

[6] We use the notation $X_p$, for a physical quantity in units of $10^p$.

[7] This conclusion is not true in general for low-mass binaries, and pulsar cavities of low-mass systems can be subject to interesting instabilities (Tavani, 1991).

## 4. Shock Radiative Regimes of Binary Pulsars

The luminosity and spectrum of high energy shock radiation depend on the various radiative timescales and their relation to the flow timescale. The flow time of particles streaming with velocity $V_f \sim c/3$ behind the nebular shock is $\tau_f \simeq r_s/V_f \simeq 50\,[1/(n_{10} f_p v_7^2)]^{1/2}\,[(c/3)/V_f]$ s. We adopt here the acceleration time $\tau_a$ of nonthermal magnetosonic acceleration of $e^{\pm}$-pairs $\tau_a \sim \Omega_i^{-1} = m_i c \gamma_1/(Z e B_2) \sim 10^2\,[\gamma_{1,6}/Z\,B_2]$ s, where $\Omega_i$ is the ion's relativistic cyclotron frequency at the shock front (HAGL92). We can assume the validity of the perpendicular shock calculations for the *apex* of the pulsar cavity. These calculations give a downstream power-law index of the $e^{\pm}$-pair momentum distribution $s \sim 2$, and a fraction $\varepsilon_a \sim 0.2$ of the flow energy flux going into the final energy of the nonthermal pairs (HAGL92). However, for a variety of binary orbital parameters, because of the rapid cooling of the shocked relativistic particles, near periastron the downstream relativistic Lorentz factor never reaches the value $\gamma_m = \gamma_1 m_i/(Z m_{\pm})$ which is obtainable only without appreciable radiative cooling. For a very eccentric orbit (as in the case of PSR 1259-63 (Johnston *et al.*, 1992) only far from periastron the downstream $\gamma$-factor can reach $\gamma_m$.

Radiative losses of the $e^{\pm}$-pairs due to synchrotron radiation in the local magnetic field at the shock and inverse Compton scattering (IC) in the background IR, optical and UV radiation field originating from the companion star surface may limit the maximum Lorentz factor attainable at the shock. This is particularly true for high-mass pulsar binaries, with luminous companion stars providing a large background of low-energy photons permeating the shock region. The synchrotron loss time is $\tau_s(\gamma) = 3 m_{\pm}^3 c^5/(e^4 B_2^2 \gamma) \sim 1.3 \cdot 10^3\,[\gamma_6 B_2^2]^{-1}$ s. The cooling time for IC scattering of high energy $e^{\pm}$-pairs against photons in the Thomson regime ($\gamma \epsilon \ll m_{\pm} c^2$, with $\epsilon$ the photon energy) is $\tau_c = (U_B/U_p)\tau_s$, where $U_B$ is the energy density of the post-shock electromagnetic field, and $U_p$ is the radiation energy density at the shock radius. The electromagnetic energy density at $r_s$ is $U_B(r_s) = B_2^2(r_s)/8\pi$, and the radiation energy density is $U_p = R_*^2 \sigma_B T^4/(\overline{R}_s^2 c)$, where $\sigma_B$ is the Stefan-Boltzmann constant, $T$ is the companion star's effective temperature, and $\overline{R}_s$ is the distance between the closest point on the surface of the companion star and the shock. The above formulae are valid only in the Thomson regime and in the case of a radiation bath at the photospheric temperature of the companion (e.g., $T \simeq 27{,}000$ K $\simeq 2.3$ eV for a typical Be star). This approximation is valid *only for* $\gamma_1 \lesssim 10^5$. However, for a typical Crab-like Lorentz factor of the relativistic particles at the shock ($\gamma_1 \sim 10^6$), Compton scattering occurs in the relativistic regime and it is necessary to take into account the Klein-Nishina decrease of the cross section as a function of center-of-mass energy. The dimensionless parameter characterizing the scattering is (Blumenthal & Gould, 1970) $\Gamma_\epsilon = 4\,\gamma\,\bar{\epsilon}/(m_{\pm} c^2)$, with $\bar{\epsilon} = 2.7\,k_B\,T$ the average photon energy of the radiation blackbody field, and $k_B$ the Boltzmann's constant. For typical Crab-like parameters of the pulsar wind and Be-star surface temperatures, $\Gamma_\epsilon \sim 50\,\gamma_{1,6}$. The Thomson approximation can be used for $\Gamma_\epsilon \lesssim 1$, but it is clearly not applicable if $\gamma_{1,6} \sim 1$. It can be shown that the ratio of the IC scattering timescale in the Klein-Nishina regime $\tau_{c,r}$ over the non-relativistic Thomson timescale $\tau_c$ in the case of PSR B1259−63 with $\gamma_1 \sim 10^6$ is $\tau_{c,r}/\tau_c \gtrsim 10$.

Different radiative regimes are produced at the shock for different combinations of timescales. A value of $\tau_a$ small compared with other relevant timescales is required to accelerate the $e^\pm$-pairs and produce a non-thermal power-law distribution of $e^\pm$ at the shock with energies $\gamma > \gamma_1$. On the other hand, the cooling processes by synchrotron and IC emission limit the non-thermal shock acceleration. The total cooling rate per radiating $e^\pm$-pair at the shock $\tau_r^{-1}$ can be written as $\tau_r^{-1} = \tau_s^{-1} + \tau_c^{-1}$. If $\tau_r$ is long compared with $\tau_f$, the radiative losses are negligible, the flow is adiabatic, and the resulting radiation nebula is *diffuse* (e.g., AT93). On the other hand, if the cooling timescale $\tau_r$ is short compared to $\tau_f$, the shock acceleration of pairs produces a spectrum of $e^\pm$ which is cut off at high energy, and a *compact* nebula forms. Finally, if $\tau_a \gg \tau_r$, no nonthermal acceleration is possible at the shock and the emerging spectrum is dominated by the cooling mechanism with the largest rate.

If heavy ions constitute a substantial component of the pulsar wind, the maximum Lorentz factor $\gamma_m^*$ that can be achieved by nonthermal acceleration of ions' magnetosonic waves at the shock is obtained from the condition $\tau_s(r_s)/\tau_a(r_s) \sim (\gamma_m^* \gamma_1)^{-1}$ which is substantially smaller than $\gamma_m$ obtained in the absence of radiative losses. Thus, different radiation regimes occur depending on the ratios of timescales $\tau_s/\tau_c, \tau_a/\tau_s, \tau_a/\tau_c$. We notice that $\tau_c$ can be much smaller than $\tau_s$ near periastron for a large variety of outflow parameters. IC scattering therefore can dominate the cooling processes at periastron unless the shock is relatively close to the pulsar. This happens for $\Upsilon \gtrsim 10^3$ and typical high-mass binary pulsar orbital parameters (TAK94). A pulsar-driven shock radiative regime dominated by IC cooling due to a companion star optical emission has not been studied previously.

4.1 INTENSITY AND SPECTRAL PROPERTIES OF SHOCK EMISSION

*4.1.1 Non-thermal 'diffuse' nebular emission.* This is the case characterized by $\tau_a \ll \tau_f \ll \tau_s, \tau_c$. This case is relevant for a fraction or whole pulsar orbits with orbital periods $P_{orb} \gtrsim 1\text{yr}$. Except for cases with $1 \lesssim \Upsilon \lesssim 10$ (with $n' = 3$, and $\sigma \gg 1$), synchrotron cooling dominates the emission for negligible emission from the companion star or large orbital distances. Since the flow time in the shock front is much less than the synchrotron loss time in the downstream flow, the flow is adiabatic both in the ions and in the pairs. Therefore, the resulting synchrotron nebula at distances from the pulsar $r \gtrsim r_s$ is diffuse. As a result, the shocked wind radiates a nebular synchrotron spectrum with emissivity $j_\epsilon \propto \epsilon^{-1/2}$ (photon energy cm$^{-3}$ s$^{-1}$). The nonthermal acceleration of pairs at the shock produces a power-law distribution of accelerated particles for the case of magnetosonic absorption of the ion's waves and the maximum Lorentz factor $\gamma_m$. Therefore, the energy spectrum is in the range $\epsilon_1 < \epsilon < \epsilon_m$, where $\epsilon_1 = 0.3 \gamma_1^2 \hbar \omega_c(B_2) \simeq (0.05\,\text{keV})\,(Z/A)^2\,(n_8\,v_6^2)^{1/2}\,(\sigma/0.005)^{1/2}$, with $\omega_c$ the electron/positron cyclotron frequency, and $\epsilon_m = \gamma_m^2 \hbar \omega_c(B_2) \simeq (210\,\text{MeV})\,(n_8\,v_6^2)^{1/2}\,(\sigma/0.005)^{1/2}$ (TAK94). Since most of the emission comes from the highest energy particles, the nonthermal part of the pair spectrum dominates the emission. For $N(\gamma) \propto \gamma^{-s}$, the bolometric luminosity of the flowing plasma is $L_w \approx V_s c n_\pm^{(nt)} \sigma_T (B_2^2/6\pi) \gamma_1^{s-1} \gamma_m^{3-s} (1-x^{3-s})/(1-x^{s-1})$, where $\sigma_T$ is the Thomson cross section, $n_\pm^{(nt)}$ the number of pairs in the nonthermal part of the pair distribution function, $V_s = (k\,r_s)^3$ (with $k \gtrsim 1$) is the volume occupied by the radiating wind downstream from the shock, and $x \equiv \gamma_1/\gamma_m = Z m_\pm/m_i$ (AT93). We estimate the total number of suprathermal pairs by noting that the shock converts a fraction $\varepsilon_a$ of the pulsar's energy loss into the energy of the nonthermal part of the pair spectrum. Therefore, it can be shown that in the downstream medium flowing with velocity $V_f$, $L_w \simeq 10^{34}\,(\sigma/0.005)\,(\varepsilon_a/0.2)\,(n_8\,v_6^2/f_p^3)^{1/2}$ erg s$^{-1}$.

*4.1.2 Non-thermal 'compact' nebular emission.* This is the case characterized by $\tau_s, \tau_c \ll \tau_f$ which applies to tight binaries or to the pulsar passage near periastron for systems similar to PSR 1259-63. We distinguish two main radiative regimes (TAK94).

*Case A*
If $\tau_a \ll \tau_f, \tau_s, \tau_c$, non-thermal acceleration of $e^\pm$-pairs at the shock is possible and the downstream power-law distribution extends from $\gamma_1$ to some $\gamma_m^*$. This case occurs only for a small $r_s$ near the pulsar and for a large ram pressure from the companion star outflow, i.e., $\Upsilon \gtrsim 100$ for high-mass binaries. Synchrotron radiation may dominate the cooling near periastron with $\tau_a \ll \tau_s$ for $\Upsilon \gtrsim 10^3$, especially if $\gamma_1 \sim 10^6$ when the relativistic IC scattering dominates. In this case, the energy spectrum is peaked at $\epsilon_1 \simeq 0.3\,\gamma_1^2\,\hbar\,\omega_c(B_2) \simeq 10$ keV and it is a power law spectrum of emissivity $j_\epsilon \sim \epsilon^{-1}$ up to an energy $\epsilon_m \sim 40$ MeV.

*Case B*
For values of $\Upsilon \lesssim 10^3$, IC scattering dominates the cooling in high-mass binaries[8] and the acceleration time $\tau_a$ becomes larger than $\tau_c$ (and $\tau_{c,r}$). In this case, the energy distribution of downstream $e^\pm$-pairs resembles a relativistic Maxwellian with no power-law tail due to shock acceleration. The optical depth $\tau^*$ for non-relativistic IC scattering in a high-mass system can be estimated as $\tau_{nr}^* \sim \sigma_T\,U_p\,r_s/\bar{\epsilon} \sim 10\,r_{s,12}$, where we used the estimate for the radiation energy density $U_p \sim (40\,\text{erg cm}^{-3})\,[(R_*/\overline{R}_s)/0.2]^2$. If IC scattering is predominantly in the Thomson regime, the photon spectrum has a Maxwellian shape with most of the energy radiated near the peak energy $E_p \sim \gamma^2 \bar{\epsilon}$. For example, in the case of PSR B1259–63, the center of mass energy in the electron frame is $\bar{\epsilon}\gamma_1/(m_\pm c^2) \sim 4.5\,\gamma_{1,6}$ and the Thomson regime applies when $\gamma_1 \lesssim 10^5$. In this case, the peak energy is $E_p \sim (45\,\text{GeV})\,\gamma_{1,5}^2$. The total luminosity of the IC spectrum is $L_{C,nr} = (4/3)\,\sigma_T\,c\,U_p\,\tilde{f}\,N\,\tilde{k}\,\gamma_1^2$, where $\tilde{k}$ is a constant of order unity, $N \sim \dot{N}\bar{\tau}$ the total number of $e^{\pm}$-pairs IC radiating per unit time, with $\dot{N} \sim \tilde{f}\,\dot{E}_R/\gamma_1\,m_{\pm}\,c^2$ the $e^{\pm}$-pair production rate, $\bar{\tau}$ the residence time $\bar{\tau} \sim r_s/(c/3)$, and $\tilde{f}$ the fraction of $N$ $e^{\pm}$-pairs near the shock. The total number of $e^{\pm}$-pairs is $N \sim 10^{38} r_{s,12}\,\gamma_{1,6}^{-1}$, and from an estimate of the solid angle of particles streaming toward the companion star, $\tilde{f} \sim 0.1$. We obtain $L_{C,nr} \sim 5 \cdot 10^{35} (\tilde{f}/0.1)\,N_{38}\,\gamma_{1,5}\,r_{s,12}$ erg s$^{-1}$. Alternately, if $\gamma_1 \sim 10^6$ as in the case of the Crab pulsar (KC84) the IC scattering is in the Klein-Nishina regime with maximum photon energy $E_{max} \sim \gamma_1\,m_{\pm}\,c^2 \simeq (500\,{\rm GeV})\,\gamma_{1,6}$. The IC luminosity is in this case, $L_{C,r} \sim (3/8)\,\sigma_T\,c\,U_p\,(m_{\pm}\,c^2/\bar{\varepsilon})^2\,\tilde{f}\,N\,[\ln(2\,\gamma_1\,\bar{\varepsilon}/m_{\pm}\,c^2) + 1/2] \sim 6 \cdot 10^{34}\,[(R_*/\overline{R}_s)/0.2]^2\,(\tilde{f}/0.1)$ erg s$^{-1}$. Note that the $e^{\pm}$-pairs of the pulsar wind can have their $\gamma_1$ lowered by IC scattering even before they reach the shock (TAK94).

---
[8]IC cooling is typically negligible in low-mass binaries (e.g., AT93).

## 5. Applications to Pulsar Systems

### 5.1 YOUNG PULSARS IN HIGH-MASS BINARIES: THE CASE OF PSR B1259-63

As binary stellar evolution predicts, young pulsars can be found orbiting around high-mass companion stars (e.g., van den Heuvel & Rappaport, 1987). The discussion of Sects. 3 and 4 applies to these systems and time variable high energy emission can be produced. The ratio $r_s/R_{lc}$ is in the range $10^4 - 10^5$, i.e., substantially smaller than the Crab Nebula value. It is therefore possible to test a region of parameter space in principle different from that one of the Crab Nebula.

The prototypical example of a powerful pulsar in a high-mass system is the PSR 1259-63 system (Johnston *et al.*, 1992). The PSR B1259−63 system provides an ideal laboratory to study the interaction of pulsar winds and circumbinary material from companion stars. PSR B1259−63 is the first radio pulsar found orbiting a massive non-degenerate stellar companion. The pulsar has spin period $P = 47.76$ ms, period derivative $\dot{P} = 2.279 \cdot 10^{-15}$, and spindown luminosity $\dot{E}_R \simeq 9 \cdot 10^{35}$ erg s$^{-1}$. The orbital period is $P_{orb} = 3.4$ yrs, and the orbital eccentricity $e = 0.87$. Its companion is SS 2883, a 10th magnitude Be star of luminosity $L_o = 5.8 \cdot 10^4\,L_\odot$, with radius $R_* \sim 11\,R_\odot \sim 8 \cdot 10^{11}$ cm and an estimated distance from Earth of 2.5 kpc. Be stars are characterized by high density mass outflows in the stellar equatorial plane and the wide orbit and high eccentricity of the PSR B1259−63 system provide a highly variable environment for the interaction between the pulsar and Be star winds. Accretion near periastron is unlikely for a broad range of reasonable outflow parameters $(\Upsilon, n')$ of the PSR B1259−63 system (TAK94).

Preliminary results of a multiwavelength campaign organized to monitor the periastron passage of PSR 1259-63 (January 9, 1994) are consistent with time variable pulsar-driven shock emission in the system. The radio pulsar disappeared around December 21, 1993 and reappeared around February 4, 1994, indicating the presence of absorbing/dispersing circumstellar material near the periastron region (Manchester, 1994). Three *ASCA* observations of this system in the energy range 0.5-10 keV revealed an *unpulsed* source with variable intensity of order of $10^{34}$ erg s$^{-1}$ and variable spectrum with photon index $\alpha \sim 1.7 - 2$ (Tavani *et al.*, 1994a; Kaspi *et al.*, 1994). The only CGRO instrument capable of detecting a faint hard X-ray source (at the level of a few mCrab) is *OSSE*. A very significant emission was detected by *OSSE* from PSR B1259−63 in the energy range 30-300 keV during a 3-week observation[9] covering the periastron passage (Grove *et al.*, 1994). A preliminary analysis gives an *OSSE* spectrum consistent with the power-law emission detected by *ASCA*. The characteristics of the X-ray and hard X-ray emission of PSR B1259−63 appear to be consistent with those expected from shock emission. The lack of pulsations of the X-ray emission (checked independently by *ASCA, BATSE* and *OSSE*) argues against accretion. Only nebular shock emission or emission due to a 'propeller effect' can explain the observations (Tavani *et al.*, 1994b). The hardness of the spectrum and its power-law nature argue in favor of nebular shock emission. We notice that the upper limits established by *COMPTEL* and *EGRET* for emission from $\sim 1$ MeV to $\sim 30$ GeV (Tavani *et al.*, 1994b) marginally exclude a shock emission mechanism dominated by IC cooling (case *B* of Sect. 4.1.2). The overall efficiency of conversion of pulsar spindown luminosity into broad band emission in the X-ray energy range is observed to be $0.01 \lesssim \varepsilon_a \lesssim 0.1$.

PSR B1259−63 is *not* a very powerful pulsar (it is about 100 time less powerful than the Crab pulsar). Furthermore, PSR B1259−63 has a long orbital period, and is rarely near its companion star. Systems which are more favorable from the point of detecting time variable nebular shock emission may yet be discovered. High-mass binaries containing Crab-like pulsars with $P_{orb} \sim 10 - 100$ days at the distance of 2-3 kpc would produce a nebular high energy flux between 10 and 100 times larger than what observed in the PSR B1259−63 system.

5.2 RECYCLED PULSARS IN LOW-MASS BINARIES

A large fraction ($\sim 50\%$) of millisecond pulsars are in low-mass binaries with white dwarf or lower main sequence companions (e.g., Batthacharya & van den Heuvel, 1991). This is in agreement with what expected from recycled old pulsars spun up during an accreting phase in low-mass X-ray binaries (e.g., Ruderman *et al.*, 1989). Even in these binaries, high energy nebular emission can be produced because of pulsar wind interaction with mass outflows from irradiated companion stars (AT93). The best example of a self-stimulated system (with shock emission providing the irradiating flux necessary to drive a mass outflow from the irradiated companion, Ruderman *et al.*, 1989) is the eclipsing millisecond pulsar PSR 1957+20 (Fruchter *et al.*, 1988). Some of these low-mass systems might contain enshrouded millisecond pulsars (Tavani, 1991) and high energy shock emission may be the only way to reveal their existence. We notice that in these systems, $r_s/R_{lc} \sim 10^3 - 10^5$, and from Eq. (4) the radius $R_G$ can be of the same order as the orbital distance.

---

[9] *OSSE* did not detect any flux from PSR B1259−63 during an observation carried out in September 1991, i.e., near apastron (Ray *et al.*, 1993).

## 6. The Nature of Galactic Gamma-Ray Sources

The unidentified $\gamma$-ray sources in the Galaxy may be *isolated* compact objects (such as $\gamma$-ray pulsars, e.g., Thompson, 1994) and *binary* systems. The $\gamma$-ray luminosity of isolated pulsars is typically in the range $10^{33} - 10^{35}$ erg s$^{-1}$. We notice that the relatively large $\gamma$-ray luminosity of the unidentified $\gamma$-sources (with $E_\gamma \gtrsim 100$ MeV, $L_\gamma \sim 10^{35} - 10^{36}$ erg s$^{-1}$ for a distance of 2-3 kpc, e.g., Kanbach, 1994) is similar to what is expected from pulsar-driven binaries. These unidentified $\gamma$-ray sources do not appear to be associated with strong accreting X-ray sources. Furthermore, several of them appear to be time variable by a factor $\sim 3$ within a timescale of weeks-months (Fichtel *et al.*, 1994). We argued here that accreting sources are *not* likely candidates for $\gamma$-ray emission. Rather, binaries containing rapidly rotating pulsars are natural candidates for high energy emission. A total of about 10 high-mass and low-mass systems with rapidly rotating pulsars are predicted within 2-3 kpc from Earth (e.g., van den Heuvel & Rappaport, 1987). The high energy shock emission is likely to be time dependent and to follow the geometry variation of the pulsar cavity as a function of orbital phase. Typical orbital periods range from tens to hundreds of days for high mass systems and from a few hours to days for low-mass systems. Monitoring galactic $\gamma$-ray sources is therefore of crucial importance for their interpretation.

At present, there is no unambiguous identification of a $\gamma$-ray source showing a modulation associated with an orbital period. The hunt for a binary signature of $\gamma$-ray sources is an important task for the CGRO instruments. An example of the work necessary to reveal the nature of $\gamma$-ray sources is the investigation of the galactic source *2CG 135+1*. Recent *EGRET* observations reveal that the new $\gamma$-ray error box of this source is consistent with the position of the radio-loud massive star *LSI 61 303* which has a distinctive periodicity of the radio continuum emission of 26.49 days (Gregory *et al.*, 1979; Taylor & Gregory, 1984). Both the intensity and spectrum of *2CG 135+1* (e.g., Kanbach, 1994; Hermsen, 1994) are consistent with pulsar-driven shock emission in a high-mass system (Tavani *et al.*, 1994c; see also Maraschi & Treves, 1981 for an earlier model). A series of *Compton* observations of this source are being carried out during 1994 to reveal a possible modulation of the signal with the $26.49^d$ period. However, a VLA pulsar search performed at different orbital phases gave a preliminary negative result[10] (Foster *et al.*, 1994). It is clear that we need more data to resolve this mystery.

Another class of interesting $\gamma$-ray sources are dense molecular clouds believed to produce $\gamma$-rays because of an enhanced target density to cosmic-ray bombardment (e.g., Bertsch *et al.*, 1993). If this interpretation is correct, the $\gamma$-ray flux should be stable in time. However, a recent analysis of *EGRET* data shows that the molecular cloud $\rho$ *Oph* (at an estimated distance $d \sim 130$ pc and spatially coincident with the $\gamma$-ray source *2CG 353+16*) contains two *time variable* point-like $\gamma$-ray sources in addition to diffuse emission (Hunter *et al.*, 1994). A radio-loud quasar (PKS 1622-253) might be the responsible for one of the two sources (Hunter *et al.*, 1994). However, the $\rho$ *Oph* cloud is a site of star formation (e.g., Wilking, 1985) and young binary systems (either containing massive stars with colliding outflows or rapidly rotating pulsars) in the *Ophiucus* cloud should also be considered as candidates for $\gamma$-ray emission. A significant contribution to the local cosmic ray flux may originate *inside* the molecular cloud (because of young pulsars) rather than outside the cloud. This may be a general phenomenon, important for the interpretation of enhanced $\gamma$-ray emission from molecular clouds. It is therefore important to test for the existence of time variable $\gamma$-ray sources in other molecular clouds. Young high-mass systems with energetic pulsars are likely to be associated with star forming regions and might be discovered as time-variable $\gamma$-ray sources. Gamma-ray sources in molecular clouds may be tracers of young pulsars in binaries. A deep pulsar search in $\rho$ *Oph* and similar clouds is also encouraged.

---

[10] Note, however, that the VLA pulsar search for *LSI 61 303* was not sensitive to a highly-dispersed Crab-like pulsar (Foster *et al.*, 1994).

The author acknowledges informative discussions and exchange of information with J. Arons, S. Digel, C. Fichtel, R. Foster, E. Grove, W. Hermsen, V. Kaspi, J. Mattox, F. Nagase and D. Thompson. Research supported by the NASA grant GRO/PFP-91-23.

# References

Arons J., Tavani M., 1993, ApJ, 403, 249 (AT93).
Bhattacharya, D. & van den Heuvel, E.P.J., 1991, Phys. Rep., 203, 1.
Bartlett, L.M., *et al.*, 1994, in *Second Compton Symposium*, eds. C.E. Fichtel, N. Gehrels, J.P. Norris, AIP Conf. no. 304, p. 67.
Bell, A.R., 1978, M.N.R.A.S., 182, 147.
Bertsch, D.L., *et al.*, 1992, Nature, 357, 306.
Bertsch, D.L., *et al.*, 1993, Ap.J., 416, 587.
Bignami, G.F. & Hermsen, W., 1983, Ann. Rev. Astron. Astrophys., 21, 67.
Blandford, R.D. & Ostriker, J.P., 1978, Ap.J., 221, L29.
Blumenthal G., Gould R., 1970, Rev. Mod. Physics, 42, 237.
Clear, J., *et al.*, 1987, A&A, 174, 85.
Cheung, W.M. & Cheng, K.S., 1993, Ap.J., 413, 697.
Coroniti, F.V., 1990, Ap.J. 349 538.
De Jager, O.C., *et al.*, 1994, in *Second Compton Symposium*, eds. C.E. Fichtel, N. Gehrels, J.P. Norris, AIP Conf. no. 304, p. 72.
Eichler, D. & Usov, V., 1993, Ap.J., 402, 271.
Fichtel, C., *et al.*, 1994, *EGRET* First Catalog, *Ap.J.Suppl.Ser.*, in press.
Foster, R.S., Tavani, M., and Frail, D., 1994, in preparation.
Fruchter, A.S., Stinebring, D.R. & Taylor, J.H., 1988, Nature, 333, 237.

Gallant, Y., Arons, J., 1994, ApJ, in press.
Gregory, P.C., *et al.*, 1979, AJ, 84 1030.
Grove, E., *et al.*, 1994, in preparation.
Halpern, J.P. & Holt, S.S., 1992, Nature, 357, 222.
Hermsen, W., 1994, these Proceedings.
Hoshino, M., Arons, J., Gallant, Y.A. & Langdon, A.B., 1992, ApJ, 390 454 (HAGL92).
Hunter, S.D., Digel, S.W., de Geus, E.J. & Kanbach, G., 1994, Ap.J., in press.
Johnston S., *et al.*, 1992, ApJ, 387, L37.
Jones, F. Baring, M. & Ellison, D., 1994, 23rd ICRC, 2, 243.
Kanbach, G., 1994, these Proceedings.
Kaspi, V., *et al.*, 1994, in preparation.
Kennel C. F., Coroniti F. V., 1984, ApJ, 283, 694.
Manchester, R.N., 1994, talk presented at the 1994 Aspen Winter Workshop, *Millisecond Pulsars: a Decade of Surprise*.
Maraschi, L., and Treves, A., 1981, M.N.R.A.S., 194, 1P.
Mészàros, P. & Rees, M.J., 1992, Ap.J., 397, 570.
Montmerle, T., 1979, Ap.J., 231, 95.
Nolan, P.L., *et al.*, 1993, Ap.J. 409 697.
Ray, P.S., *et al.*, 1993, in *Second Compton Symposium*, eds. C.E. Fichtel, N. Gehrels, J.P. Norris, AIP Conf. no. 304, p. 249.
Rees, M.J. & Gunn, J.E., 1974, M.N.R.A.S., 167, 1.
Ruderman, M., Shaham, J., and Tavani, M., 1989, Ap.J., 336, 507.
Swanenburg, B.N., *et al.*, 1981, Ap.J.Letters, 243, L69.
Tavani, M., 1991, Ap.J.Letters, 379 L69.
Tavani, M., Arons, J., Kaspi, V., 1994, Ap.J.Letters, in press (TAK94).
Tavani, M., *et al.*, 1994a, in preparation.
Tavani, M., *et al.*, 1994b, in preparation.
Tavani, M., *et al.*, 1994c, in preparation.
Tavani, M., 1994d, Ap.J.Letters, in press.
Taylor, A. R. and Gregory, P. C. 1984, Ap.J., 283, 273.
Thompson, D.J., 1994, in *Second Compton Symposium*, eds. C.E. Fichtel, N. Gehrels, J.P. Norris, AIP Conf. no. 304, p. 57.
Vacanti, G., *et al.*, 1991, Ap.J. 377 467.
van den Heuvel, E.P.J. & Rappaport, S.A., 1987, in *Physics of Be Stars*, eds. A. Slettbak & T.P. Snow (Cambridge: Cambridge Un. Press), p. 291.
Waters L. B. F. M., *et al.*, 1988, A&A, 198, 200.
Wilking, B.A., 1985, in *Nearby Molecular Clouds*, ed. G. Serra (Berlin, Springer Verlag), p. 104.
Williams, P.M., *et al.*, 1990, M.N.R.A.S., 243, 662.